\documentclass{elsart}
\usepackage{amssymb}
\usepackage{graphicx}
\usepackage{amsmath}

\begin{document}
\newcommand{\decayarrow}{\makebox[0mm][l]{\rule{0.33em}{0mm}\rule[0.55ex]{0.044em}{1.55ex}}\rightarrow}
\begin{frontmatter}

\title{Search for $^{~6}_{~\Lambda}$H and $^{~7}_{~\Lambda}$H with the 
  (K$^{-}_{stop}$,$\pi^{+}$) reaction}

\centering{FINUDA Collaboration}

\author[polito]{M.~Agnello},
\author[victoria]{G.~Beer},
\author[lnf]{L.~Benussi},
\author[lnf]{M.~Bertani},
\author[korea]{H.C.~Bhang},
\author[lnf]{S.~Bianco},
\author[unibs]{G.~Bonomi},
\author[unitos]{E.~Botta},
\author[units]{M.~Bregant},
\author[unitos]{T.~Bressani},
\author[unitos]{S.~Bufalino},
\author[unitog]{L.~Busso},
\author[infnto]{D.~Calvo},
\author[units]{P.~Camerini},
\author[infnto]{P.~Cerello},
\author[uniba]{B.~Dalena\thanksref{corresponding}},
\author[unitos]{F.~De~Mori},
\author[uniba]{G.~D'Erasmo},
\author[uniba]{D.~Di~Santo},
\author[infnba]{D.~Elia},
\author[lnf]{F.~L.~Fabbri},
\author[unitog]{D.~Faso},
\author[infnto]{A.~Feliciello},
\author[infnto]{A.~Filippi},
\author[infnpv]{V.~Filippini\thanksref{deceased}},
\author[infnba]{R.A.~Fini},
\author[uniba]{E.~M.~Fiore},
\author[tokyo]{H.~Fujioka},
\author[lnf]{P.~Gianotti},
\author[infnts]{N.~Grion},
\author[jinr]{A.~Krasnoperov},
\author[infnba]{V.~Lenti},
\author[lnf]{V.~Lucherini},
\author[infnba]{V.~Manzari},
\author[unitos]{S.~Marcello},
\author[tokyo]{T.~Maruta},
\author[teheran]{N.~Mirfakhrai},
\author[cnr]{O.~Morra},
\author[kek]{T.~Nagae},
\author[riken]{H.~Outa},
\author[lnf]{E.~Pace},
\author[lnf]{M.~Pallotta},
\author[uniba]{M.~Palomba}, 
\author[infnba]{A.~Pantaleo},
\author[infnpv]{A.~Panzarasa},
\author[infnba]{V.~Paticchio},
\author[infnts]{S.~Piano},
\author[lnf]{F.~Pompili},
\author[units]{R.~Rui},
\author[uniba]{G.~Simonetti},
\author[korea]{H.~So},
\author[jinr]{V.~Tereshchenko},
\author[lnf]{S.~Tomassini},
\author[kek]{A.~Toyoda},
\author[infnto]{R.~Wheadon},
\author[unibs]{A.~Zenoni}

\thanks[corresponding]{corresponding author. e-mails: dalena@ba.infn.it; fax:+39.080.5443151}
\thanks[deceased]{This paper is dedicated to the memory of our colleague and
  friend Valerio Filippini}
\address[polito]{Dip. di Fisica Politecnico di Torino, via Duca degli Abruzzi
Torino, Italy, and INFN Sez. di Torino, via P. Giuria 1 Torino, Italy}
\address[victoria]{University of Victoria, Finnerty Rd.,Victoria, Canada}
\address[lnf]{Laboratori Nazionali di Frascati dell'INFN, via E. Fermi 40
Frascati, Italy}
\address[korea]{Dep. of Physics,
Seoul National Univ., 151-742 Seoul, South Korea}
\address[unibs]{Dip. di Meccanica, Universit\`a di Brescia, via Valotti 9 Brescia, Italy and INFN Sez. di Pavia, via Bassi 6 Pavia, Italy}
\address[unitos]{Dipartimento di Fisica Sperimentale, Universit\`a di
Torino, via P. Giuria 1 Torino, Italy, and INFN Sez. di Torino,
via P. Giuria 1 Torino, Italy}
\address[units]{Dip. di Fisica Univ. di Trieste, via Valerio 2 Trieste, Italy and INFN, Sez. di Trieste, via Valerio 2 Trieste, Italy}
\address[unitog]{Dipartimento di Fisica Generale, Universit\`a di
Torino, via P. Giuria 1 Torino, Italy, and INFN Sez. di Torino,
via P. Giuria 1 Torino, Italy}
\address[infnto]{INFN Sez. di Torino, via P. Giuria 1 Torino, Italy}
\address[uniba]{Dip. InterAteneo di Fisica, via Amendola 173 Bari, Italy and INFN Sez. di Bari, via Amendola 173 Bari, Italy }
\address[infnba]{INFN Sez. di Bari, via Amendola 173 Bari, Italy }
\address[infnpv]{INFN Sez. di Pavia, via Bassi 6 Pavia, Italy}
\address[tokyo]{Dep. of Physics Univ. of Tokyo, Bunkyo Tokyo 113-0033, Japan}
\address[infnts]{INFN, Sez. di Trieste, via Valerio 2 Trieste, Italy}
\address[jinr]{JINR, Dubna, Moscow region, Russia}
\address[teheran]{Dep of Physics Shahid Behesty Univ., 19834 Teheran, Iran}
\address[cnr]{INAF-IFSI Sez. di Torino, C.so Fiume, Torino, Italy
and INFN Sez. di Torino,
via P. Giuria 1 Torino, Italy}
\address[kek]{
High Energy Accelerator Research Organization (KEK), Tsukuba, Ibaraki
305-0801, Japan}
\address[riken]{RIKEN, Wako, Saitama 351-0198, Japan}

\begin{abstract}
The production of neutron rich $\Lambda$-hypernuclei via the 
(K$^{-}_{stop}$,$\pi^{+}$) reaction has been studied using data collected with
the FINUDA spectrometer at the DA$\Phi$NE $\phi$-factory (LNF). The analysis of
the inclusive $\pi^{+}$ momentum spectra is presented and an upper limit for the 
production of $^{~6}_{~\Lambda}$H and $^{~7}_{~\Lambda}$H from 
$^{6}$Li and $^{7}$Li, is assessed for the first time.


\end{abstract}

\end{frontmatter}

\section{Introduction} \label{par1.0}

As pointed out by Majling~\cite{Maji95}, 
$\Lambda$-hypernuclei may be even better candidates than ordinary nuclei
to exhibit unusually large values of $N/Z$ and halo phenomena.
In fact, a $\Lambda$-hypernucleus is more stable than an ordinary nucleus
due to the compression of the 
nuclear core and to the addition of extra binding energy from the $\Lambda$
hyperon (playing the so called ``glue-like role of the $\Lambda$'')~\cite{Tret01}.
From the 
hypernuclear physics point of view, the attempt to extend our knowledge towards
the limits of nuclear stability, exploring strange systems with high $N/Z$ ratio, can
provide more information both on baryon-baryon interactions and on the behavior
of hyperons in a medium with much lower density than ordinary $\Lambda$-hypernuclei. 
Furthermore, the role of the three-body $\Lambda$$NN$ force related to the 
``coherent \mbox{$\Lambda$-$\Sigma$} coupling'' has connections with nuclear 
astrophysics~\cite{Yama01}, as previously proposed in theoretical
calculations of high density nuclear matter (neutron stars)~[4-5].
In particular, there is great interest in the possible existence of
$^{~6}_{~\Lambda}$H; in fact, theoretical calculations predict the existence of a 
stable single-particle state with a binding energy of 5.8 MeV from the 
\mbox{$^{5}\rm{H} + \Lambda$} threshold (+1.7 MeV~\cite{HaevH5}), when the 
\mbox{$\Lambda$-$\Sigma$} coupling term is considered.
Without this coupling force the state would be very close to the
\mbox{$^{~4}_{~\Lambda}$H $+ 2n$} threshold~[7-8].\\
Experimentally, the production of neutron rich $\Lambda$-hypernuclei is more 
difficult than standard $\Lambda$-hypernuclei, whose one-step
direct production reactions, such as ($K^{-}$, $\pi^{-}$) and ($\pi^{+}$, $K^{+}$),
access only a limited region in the hypernuclear chart, rather close to the stability
region. 
On the contrary, neutron rich $\Lambda$-hypernuclei can be produced by means of different 
reactions based on the double charge-exchange (DCX) mechanism, such as  
($\pi^{-}$, $K^{+}$) and ($K^{-}$, $\pi^{+}$).
The latter reaction, studied in this article, proceeds through the following two
elementary reactions:
\begin{center}
\begin{eqnarray}\label{twost}
  K^- + p \rightarrow \Lambda + \pi^0;&\quad \pi^0 + p \rightarrow n + \pi^+\\
  \nonumber\\
  K^- + p \rightarrow \Sigma^- + \pi^+;&\quad \Sigma^-~p \leftrightarrow 
  \Lambda~n  
\end{eqnarray}
\end{center}
Process (1) is a two step reaction in which a strangeness exchange
is followed by a pion charge exchange. Process (2) is a single step 
reaction with a $\Sigma^{-}$ admixture (due to the 
\mbox{$\Sigma^-~p \leftrightarrow \Lambda~n$} coupling~\cite{Tetry03}). 
Owing to these features, both processes usually have lower cross sections than
one step reactions.\\
The first experimental attempt to produce neutron rich $\Lambda$-hypernuclei via the 
($K^{-}_{stop}$,$\pi^{+}$) reaction was carried out at KEK~\cite{kub96}. 
An upper production limit (per stopped kaon) was obtained 
for $^{12}_{~\Lambda}$Be, $^{~9}_{~\Lambda}$He and $^{16}_{~\Lambda}$C 
hypernuclei. The results are in the range \mbox{(0.6$\div$2) $\times$ 10$^{-4}$};
note that the theoretical values calculated
by T.Yu. Tetryakova and D.E. Lanskoy~\cite{Tetry03} on $^{12}_{~\Lambda}$Be and
$^{16}_{~\Lambda}$C are in the range (10$^{-6}$ $\div$ 10$^{-7}$) per stopped
kaon, {\it i.e.} at least one order of magnitude less than the experimental 
upper limits and three orders smaller than the usual ($K^{-}_{stop}$, $\pi^{-}$) 
one-step reaction rates on the same targets (\mbox{10$^{-3}$}).
Recently, a KEK experiment~\cite{Saha05} claimed to have observed the production of
$^{10}_{~\Lambda}$Li in the ($\pi^{-}$, $K^{+}$) reaction on a $^{10}$B target. 
The published results are not directly comparable with theoretical predictions,
since no discrete structure was observed and the production cross section has been
integrated over the whole bound region \mbox{(0 $<$ $B_{\Lambda}$ $<$ 20 MeV)}. 
Furthermore, the experimental trend of the cross section energy dependence
strongly disagrees with theoretical predictions~\cite{Lansk05}.\\
This circumstance has stimulated a renewed interest in neutron rich 
$\Lambda$-hypernuclei, in particular in $^{~6}_{~\Lambda}$H and $^{~7}_{~\Lambda}$H.
Their production rates have no theoretical predictions nor experimental measurements.\\
The present paper shows results concerning these neutron rich $\Lambda$-hypernuclei
studied in the FINUDA experiment, where they can be produced through
processes (1) and (2) with a $K^{-}$ at rest. In both cases a final state with a $\pi^{+}$
and a $\Lambda$-hypernucleus is produced. The overall production reaction is:
\begin{center}
  \begin{eqnarray}\label{twost}
    K^{-}_{stop} + ^{A}(Z) \rightarrow ^{A}_{~\Lambda}(Z-2) + \pi^{+}.
  \end{eqnarray}
\end{center}
The residual $\Lambda$-hypernucleus has two protons less and one neutron more
than the target nucleus.\\
In the following, after a short description of the FINUDA experimental apparatus, the
analysis of the $\pi^{+}$ inclusive momentum spectra is presented and discussed.

\section{The FINUDA spectrometer}

A detailed description of the apparatus can be found in Refs.~[13-14-15].
We recall here only some features, mainly related to the present analysis. The
main goal of the experiment is to study the formation and the decay properties
of $\Lambda$-hypernuclei produced by the strangeness exchange reaction
($K^{-}_{stop}$, $\pi^{-}$). The $K^{-}$ comes from the decay at rest of the $\phi$
produced by DA$\Phi$NE. Its low kinetic energy allows it to be slowed down and stopped
in thin targets ($0.21-0.38$ g/cm$^2$).
Thus prompt $\pi^{-}$'s emitted after hypernuclei formation are 
minimally degraded and their momentum is measured with a small uncertainty 
(0.6$\%$ FWHM choosing high quality tracks). This allows the determination of the
hypernuclear levels with a resolution of $\sim$ 1 MeV FWHM.\\
\begin{figure}[!h]
    \centering
    \includegraphics[width = 11.cm, angle = 270]{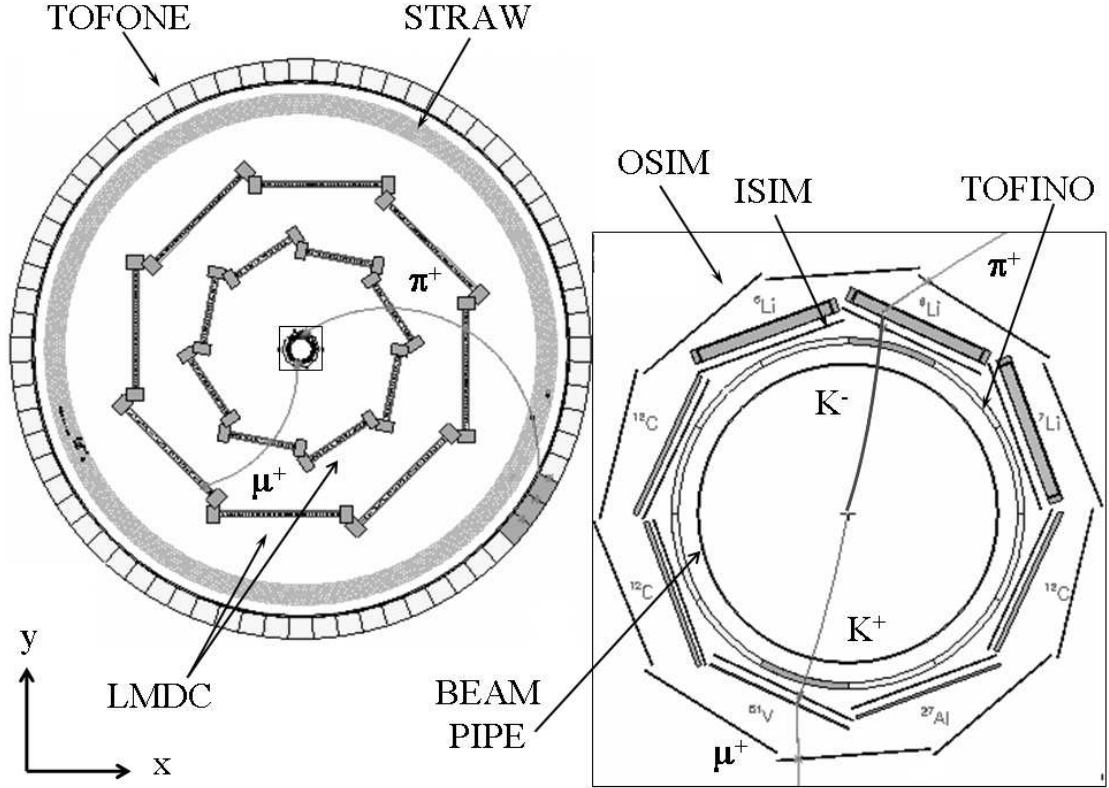}
    \caption{\small Picture of a $\pi^{+}$, coming from a $K^{-}$ stopped in a 
      $^{6}$Li target, recorded by the FINUDA tracking region
      (positive tracks turn clockwise). The track coming from
      the $K^{+}$ stopping point is a $\mu^{+}$ from $K_{\mu2}$ decay.
      In the inset, the ($K^{+}$,$K^{-}$) pair from $\phi$ decay recorded by
      the FINUDA vertex detector and the eight targets employed can be seen.
       \label{fv}}
\end{figure}
The ($K^+K^-$) pairs are detected by a barrel of 12 scintillator slabs
(TOFINO) 2.3 mm thick and 20 cm long, surrounding the beam pipe. This detector provides
fast trigger signals and, together with the external scintillator barrel (TOFONE), 
measures the Time Of Flight (TOF) of charged and neutral particles, with
a resolution $\sigma_{TOF}$(TOFONE - TOFINO) $\sim$ 420 ps.  
The TOFINO barrel is surrounded by an octagonal Inner array of SIlicon
Micro-strip detectors (ISIM). This detector is used to identify the
($K^+K^-$) pairs with a $\Delta$$E$/$\Delta$$x$
resolution of 20$\%$ and to determine 
their interaction points in the targets with a resolution of a few hundred
microns (mainly due to multiple scattering).
Eight targets surround the ISIM modules, as shown in Fig.~\ref{fv}. Five
different target materials were used during the first data
taking: two targets of $^{6}$Li, one of $^{7}$Li,
three of $^{12}$C, one of $^{27}$Al and one of $^{51}$V. 
Charged particle tracks coming from the targets are measured by an Outer array
of ten double-sided SIlicon Micro-strip detector modules (OSIM), two arrays of
eight planar Low-Mass Drift Chambers (LMDC), immersed in a He 
atmosphere to reduce Coulomb multiple scattering, and a straw tube detector
assembly (STRAW), composed by six layers of longitudinal and stereo tubes. The crossing
point of the incident particles can be extrapolated using the information of the
fired tubes, with a spatial resolution $\sigma_{z}$~$\sim$~500~$\mu$m and 
$\sigma_{\rho\phi}$~$\sim$~150~$\mu$m.\\
The external time of flight detector barrel (TOFONE) is composed of 72 
scintillator slabs, 10 cm thick and 255 cm long, providing fast signals to 
the trigger and TOF.\\
Particle identification of the track is allowed by the energy loss $\Delta$$E$/$\Delta$$x$
in OSIM and the TOF. Note that the TOF evaluated between TOFINO and TOFONE include,
for tracks emerging from a target and reaching TOFONE, the negligible contribution
of ~200 ps due to the $K^{-}$ time of flight from TOFINO to its stopping point in the target,
well within TOF timing resolution.\\
A sample of data, corresponding to an integrated luminosity of about 190
pb$^{-1}$, has been collected during the first FINUDA data taking.

\section{Analysis of $\pi^{+}$ inclusive momentum distributions}
\label{analisi}

\begin{figure}
 \centering
  \includegraphics[height=24.pc]{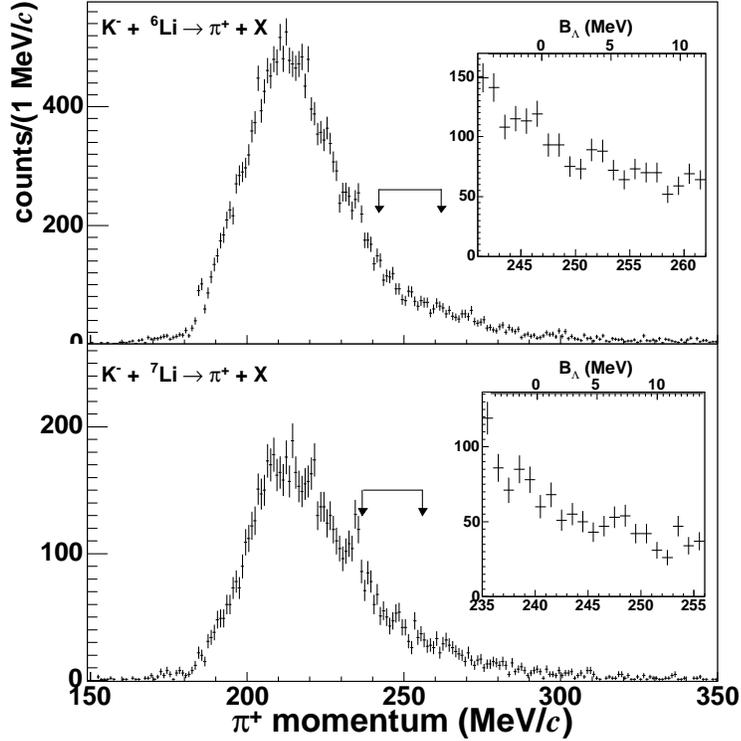}
  \caption{\small Inclusive $\pi^+$ momentum spectra.
    Expanded views of the regions between the two arrows 
    are shown in the insets, with the corresponding $\Lambda$ binding energy values on
    top.}
  \label{finpi}
\end{figure}

\begin{figure}
 \centering
  \includegraphics[height=24.pc]{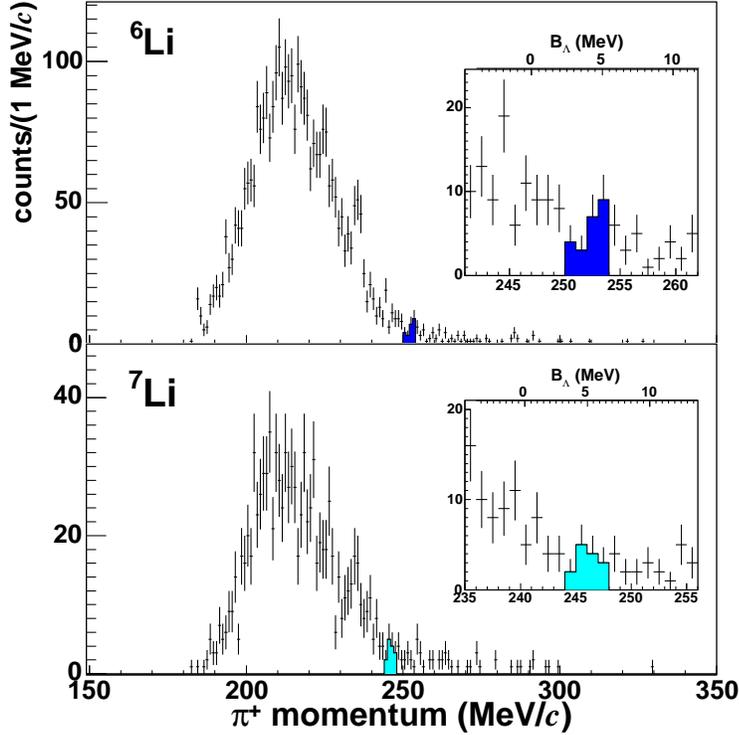}
  \caption{\small Inclusive $\pi^+$ momentum spectra after the background reduction
    carried out as described in the text. ROIs are highlighted and enlarged
    views of the same spectra around the ROIs, with the $\Lambda$ binding
    energy axis, are shown in the insets.}
  \label{finhist}
\end{figure}

The data used for this analysis (\mbox{$\sim$ 4 $\times$ 10$^{6}$} 
$K^{-}_{stop}$ events) refer to the lighter targets, 
namely two of $^{6}$Li and one of $^{7}$Li, where the two elementary processes (1)
and (2) lead to the formation of the following hypernuclei:
\begin{center}
  \begin{eqnarray}\label{global}
    & K^{-}_{stop} + ^{6}\rm{Li} \rightarrow ^{~6}_{~\Lambda}\rm{H} + \pi^+\\
    & K^{-}_{stop} + ^{7}\rm{Li} \rightarrow ^{~7}_{~\Lambda}\rm{H} + \pi^+
  \end{eqnarray}
\end{center}
The emitted $\pi^+$ momenta are related to the $\Lambda$ binding energies $B_{\Lambda}$
of the predicted hypernuclear ground state of $^{~6}_{~\Lambda}$H 
(B$_{\Lambda}$= 4.1 MeV~\cite{Myint00}) and of $^{~7}_{~\Lambda}$H 
(B$_{\Lambda}$= 5.2 MeV~\cite{Maji95}) through momentum
and energy conservation, and are evaluated as $\sim$ 252 and $\sim$ 246 MeV/$c$ respectively. 
The candidate events were selected by requiring a successfully reconstructed positive
track associated with a $K^{-}$ stopped in the selected target. The positive particle
associated with this track is identified as a $\pi^{+}$ by means of its 
$\Delta$$E$/$\Delta$$x$ and of its TOF (``{\em soft }'' cuts).
The momentum spectra of the selected $\pi^+$ are shown in Fig.~\ref{finpi}. The spectra
are not corrected for acceptance. This influences their shape mainly 
in the momentum region 180-220 MeV/$c$, due to the kinematic cut of the spectrometer.
In the same figure a residual 236 MeV/$c$ peak, due to $K_{\mu2}$ decay contamination,
coming from a few $K^{+}$/$K^{-}$ misidentified events and not completely removed by TOF selection,
can be seen. No significant structures are observed in the B$_{\Lambda}$ region 
(0 $<$ B$_{\Lambda}$ $<$ 10 MeV) as can be seen in the inset of Fig.~\ref{finpi}.\\
The bulk of the spectrum is due to $\pi^+$ coming from $\Sigma^+$ decay, 
produced in the following two quasi-free reactions~\cite{Ohni97}:
\begin{eqnarray}\label{rnppiu}
  K^{-} + p \to & \Sigma^{+}& +~~ \pi^{-} \\ 
  & \decayarrow & n + \pi^{+} \qquad (\sim130 < p_{\pi^{+}}< \sim250~\rm{MeV}/{\it c})\nonumber
\end{eqnarray}
\begin{eqnarray}\label{rnppppiu}
  K^{-} + pp \to & \Sigma^{+}& +~~ n  \\ 
  &  \decayarrow & n + \pi^{+} \qquad (\sim100 < p_{\pi^{+}}< \sim320~\rm{MeV}/{\it c}). \nonumber 
\end{eqnarray}
In particular, the $\pi^{+}$ counts in the momentum region of interest are mostly due to reaction
(\ref{rnppppiu}), to some $K_{\mu2}$ in-flight decay contamination and a small contribution
from the high momentum tail of reaction (\ref{rnppiu}).\\
In order to reduce the contribution of the above events to the $\pi^{+}$ counts 
in the momentum region of interest,
further event selections can be applied, taking advantage of the tracking
capabilities of the FINUDA spectrometer. To this purpose, we focused on
the distance between the $K^{-}$ absorption point and the $\pi^{+}$
origin point, estimated by the reconstruction algorithm as the point of
closest approach between the two extrapolated 
tracks beyond ISIM and back from OSIM respectively, towards a plane in the 
target volume. A cut on the value of this distance can reduce the
contribution from in-flight $\Sigma^+$ decay of reactions (6) and (7) and from in-flight
contamination; in fact, the $\pi^{+}$ or $\mu^{+}$ coming from these decays
can be reconstructed some millimeters apart from the
$K^{-}$ stopping point (whereas, the $\pi^{+}$ following the hypernuclear
formation is produced at the same point in which the $K^{-}$ is absorbed
at rest). Using two distinct simulations, one for the background and one for the signal,
a 2 mm cut (in the following referred to as ``{\em hard }'' cut)
in such a distance selects almost 50$\%$ of pions coming from the hypernuclear formation
and 10$\%$ of background. Therefore, this selection improves the
signal-to-noise ratio by a factor $\sim$ 5.
Applying this cut, any contributions in the high momentum tail due to in-flight decays,
is greatly reduced (see Fig.~\ref{finhist}, to be compared with Fig.~\ref{finpi}).
The $K_{\mu2}$ contamination at 236 MeV/$c$ produced in the misidentified vertex
is not affected by the ``{\em hard }'' cut, as expected.
From the inset of Fig.~\ref{finhist} for $^{6}$Li, there is an indication for a peak at $\sim$
254 MeV/$c$, corresponding to a B$_{\Lambda}$ of $\sim$ 5.6 MeV. We studied the statistical
significance for such a signal with three different hypothesis on the background,
due to the reactions (\ref{rnppiu}) and (\ref{rnppppiu}), that has been parametrized as in 
\cite{Carb05} but we found that the C.L. is $<$ 90$\%$.
Therefore an upper limit at 90$\%$ C.L. for
the experimental production rate of the neutron rich $\Lambda$-hypernuclei $^{~6}_{~\Lambda}$H and 
$^{~7}_{~\Lambda}$H is evaluated.\\
To this purpose the maximum number of $\pi^{+}$ counts to be ascribed
to neutron rich $\Lambda$-hypernuclei formation has to be estimated.
Therefore a Region Of Interest (ROI) is defined in each spectrum, centered
at the $\pi^{+}$ momentum value corresponding to the predicted B$_{\Lambda}$. 
The ROI widths have been set to $\pm$ 2 $\sigma_{p}$, where 
$\sigma_{p}$ is the standard deviation of the peak momentum resolution (0.9$\%$ FWHM).
This value is estimated using the monochromatic $\mu^{+}$ peak at 236 MeV/$c$
using the track selection conditions of the present analysis. We do not require
high quality tracks, in order to have as much statistics as possible to
study such rare events.

\section{Production rates and upper limit evaluation}
 \label{rate}

The $\pi^{+}$ production rate $R$ per stopped $K^{-}$ is given by the ratio
of the number of the $\pi^{+}$ produced by any concurrent reaction following the
$K^{-}$ stopping in the target and the number of the stopped $K^{-}$ ($N(K^{-}_{stop})$)
considered in the analysis. 
The number of produced $\pi^{+}$ is given by the number ($N_{\pi^+}$) of measured $\pi^{+}$
weighted by the intrinsic detector efficiency $\varepsilon_{D}$($\pi^+$)
and the global efficiency $\varepsilon_{G}$($\pi^+$) in the ROI. Hence:
\begin{equation}\label{count1}
  R = \frac{N_{\pi^+}}{N(K^{-}_{stop}) \cdot \varepsilon_{D}(\pi^+)\cdot 
    \varepsilon_{G}(\pi^+)}
\end{equation}
where $\varepsilon_{G}$($\pi^+$) takes
into account the trigger efficiency, the apparatus
geometrical acceptance and the efficiency of the reconstruction algorithm.

Instead of measuring directly these efficiencies we exploit the FINUDA unique
feature of being able to detect back-to-back ($K^-$,$K^+$) pairs, to relate them
to the $\mu^{+}$ particles as follows. 
The number of measured $\mu^+$ ($N_{\mu^+}$) coming from $K_{\mu2}$
decay and emerging from the same target ({\it i.e.} a different sample of data)
can be entered in the following Branching Ratio definition:
\begin{equation}\label{count2}
  BR(K_{\mu2})=\frac{N_{\mu^+}}{N(K^{+}_{stop})\cdot \varepsilon_{D}(\mu^+)
  \cdot \varepsilon_{G}(\mu^{+})}  
\end{equation}
where $N(K^{+}_{stop})$ is the number of $K^{+}$ stopped in the same 
target, $\varepsilon_{D}$($\mu^+$) the intrinsic detector efficiency and 
$\varepsilon_{G}$($\mu^+$) the global $\mu^{+}$ efficiency.
The rate $R$ is then estimated in terms of the known Branching Ratio of the
$K_{\mu2}$ decay process \mbox{($BR(K_{\mu2}$)= 0.6343)} using the following
formula (folding (8) and (9)):
\begin{equation}\label{rate1}
  R = \frac{N_{\pi^+}}{N_{\mu^+}} \cdot \frac{N(K^{+}_{stop})}{N(K^{-}_{stop})}
  \cdot \frac{\varepsilon_{D}(\mu^+)}{\varepsilon_{D}(\pi^+)} \cdot
  \frac{\varepsilon_{G}(\mu^+)}{\varepsilon_{G}(\pi^+)} \cdot  BR(K_{\mu2})
\end{equation}
It is reasonable to assume that $\mu^+$ and $\pi^+$ from each target cross
the same section of the FINUDA apparatus and then the terms $\varepsilon_{D}$($\pi^{+}$)
and $\varepsilon_{D}$($\mu^{+}$) cancel out,
whereas the $\varepsilon_{G}$ factors for the two different particles
can be estimated by means of the FINUDA Monte Carlo simulation program~\cite{Zeno04}.\\
In order to evaluate the Upper Limit (U.L.) from the $\pi^{+}$ counting rate,
this has to be scaled down by a statistical factor which
estimates the maximum fraction of the $\pi^{+}$ counts that may be ascribed to
neutron rich $\Lambda$-hypernuclei formation, within the considered
Confidence Level (C.L.). 
We consider the total number of counts in the ROI as the sum of the signal counts $S$ and 
of the background counts $B$ ($N_{\pi^{+}} = S + B$). The statistical factor is defined
as $S/N_{\pi^{+}}$.
In addition, since the sample of 
$N_{\pi^{+}}$ as well as $S$ and $B$ obey Poisson statistics, the minimum value of $B$
compatible with the fluctuations, within the fixed C.L., can be evaluated
numerically from the following integral equations:  
\begin{equation}
  \left\{ \begin{array}{ll} 
    \int^{\infty}_{N_{C.L.}} \frac{\mu^{N_{\pi^+}}}{N_{\pi^+}!} e^{-\mu} 
    d\mu &= C.L. \\ [.4cm]
    \int_{0}^{B_{C.L.}} \frac{\mu^{B}}{B!} e^{-\mu} 
    d\mu &= C.L.
  \end{array} \right. 
\end{equation}
The minimum $B$ value satisfying the condition $B_{C.L.} \ge N_{C.L.}$
is computed with an iterative procedure. 
Table~\ref{finval} shows the U.L. values of production rate per
stopped kaon at a 90$\%$ C.L. for 
\mbox{$^{6}$Li$(K^{-}_{stop},\pi^{+})^{6}_{~\Lambda}$H} and  
\mbox{$^{7}$Li$(K^{-}_{stop},\pi^{+})^{7}_{~\Lambda}$H} reactions.
The U.L. values corresponding to the ROI's shifted at $\pm$ 1 MeV/$c$
do not show great variations with respect to previous ones (last columns 
in Table~\ref{finval}).
\begin{table}
  \begin{center} 
    \begin{tabular}{| c c c c c| }\hline 
      \textbf{Target} & \textbf{$\Lambda$-Hyp} & \textbf{U.L.} (90$\%$ C.L.)
      & $\Delta$(\textbf{U.L.}) & $\Delta$(\textbf{U.L.}) \\
      & & & @ +1 MeV/$c$ & @ -1 MeV/$c$\\
      \hline \hline
      $^{6}$Li & $^{~6}_{~\Lambda}$H & (2.5$\pm$0.4$_{stat}$
      $^{+0.4}_{-0.1}$$_{syst}$)
      $\times$ 10$^{-5}$& -0.4 $\times$ 10$^{-5}$ & +0.0 $\times$ 10$^{-5}$\\
      \hline
      $^{7}$Li & $^{~7}_{~\Lambda}$H & (4.5$\pm$0.9$_{stat}$
      $^{+0.4}_{-0.1}$$_{syst}$)
      $\times$ 10$^{-5}$& -0.5 $\times$ 10$^{-5}$ & +0.1 $\times$ 10$^{-5}$\\
      \hline
    \end{tabular}
    \caption{\small Upper Limits (U.L.) at a 90$\%$ C.L. of neutron rich 
      $\Lambda$-hypernuclei production rate per stopped $K^{-}$ for the 
      ($K^{-}_{stop},\pi^+$) reaction on the two light target nuclei considered
      in the analysis. The last two columns represent the variations in the U.L. values
      obtained shifting the center of the ROI at $\pm$ 1 MeV/$c$, respectively.}
    \label{finval}
  \end{center} 
\end{table}
\\
It is worth mentioning that the same procedures, described in this paper,
have been applied to the three $^{12}$C targets. There is no evidence of $^{12}_{~\Lambda}$Be
production and the upper production rate limit is 
(2.0$\pm$0.4$_{stat}$$^{+0.3}_{-0.1}$$_{syst}$) $\times$10$^{-5}$/K$^{-}_{stop}$ which
improves the value published by Kubota {\em et al.}~\cite{kub96} that is  
6.1 $\times$ 10$^{-5}$/K$^{-}_{stop}$. The U.L. values evaluated for the ROI center
shifted at $\pm$ 1 MeV/$c$ do not show any significant variations with respect to
the previous one.      

\section{Conclusions}

In order to investigate the production of neutron rich $\Lambda$-hypernuclei,
events with a $\pi^+$ in the final state have been selected and analyzed after
the first FINUDA data taking.
In the analysis presented here, we have applied either only ``{\em soft }'' cuts
on the $\pi^+$ momentum, with the aim of identifying good $\pi^+$ while retaining as many
events as possible, or the same plus an additional ``{\em hard }'' cut, designed to
improve the signal-to-noise ratio within the chosen ROIs. 
None of the approaches showed any significant structure at minimum 90$\%$ C.L. either 
in the ROIs or in the whole $\Lambda$ bound region. 
Upper Limits for the production rate of $^{~6}_{~\Lambda}$H  and $^{~7}_{~\Lambda}$H at 90$\%$
C.L., in ($K^{-}_{stop}, \pi^+$) reactions, are (2.5$\pm$0.4$_{stat}$$^{+0.4}_{-0.1}$$_{syst}$) 
$\times$ 10$^{-5}$/K$^{-}_{stop}$ and (4.5$\pm$0.9$_{stat}$$^{+0.4}_{-0.1}$$_{syst}$)
$\times$ 10$^{-5}$/K$^{-}_{stop}$ respectively.

\section{Acknowledgments}
The authors thank Prof. I. Lazzizzera for the statistical data analysis support
of this paper, Dr. T.Yu. Tetryakova and Prof. D.E. Lanskoy for discussions, the 
DA$\Phi$NE staff for their skillful handling of the
collider and the FINUDA technical staff for their continuous support.


\end{document}